\documentclass[aps,preprint,amssymb,12pt,floatfix]{revtex4}

\usepackage{graphicx}
\usepackage{dcolumn}
\usepackage{bm}
\usepackage{epstopdf}
\usepackage{textcomp}
\usepackage{amsmath}
\usepackage[usenames,dvipsnames]{xcolor}

\begin{document}

\title{Force-Induced Unzipping Transitions in an Athermal Crowded Environment}
\author{David L. Pincus}
\author{D. Thirumalai}
\email{thirum@umd.edu}
\affiliation{Institute for Physical Science and Technology, University
of Maryland, College Park, Maryland 20742}
\date{\today}

\begin{abstract}
Using  theoretical arguments and extensive Monte Carlo (MC) simulations
of a coarse-grained three-dimensional off-lattice model of a $\beta $-hairpin, we 
demonstrate that the equilibrium critical force, $F_{c}$, needed to unfold the biopolymer 
increases  non-linearly with increasing volume fraction occupied by the spherical macromolecular crowding agent.  
Both scaling arguments and MC simulations show  that the critical force  increases as $F_{c} \approx \varphi _{c}^{\alpha}$. The  exponent $\alpha$ is linked to the Flory exponent relating the size of the unfolded state of the biopolymer and the number of amino acids. The predicted power law dependence is confirmed in simulations of the dependence of the isothermal extensibility and the fraction of native contacts on   $\varphi _{c}$. 
We also show using MC simulations that $F_{c}$ is  linearly dependent on the average osmotic 
pressure ($\mathrm{P}$) exerted by the crowding agents  on the $\beta$-hairpin.  The highly significant linear correlation coefficient
of 0.99657 between $F_{c}$ and $\mathrm{P}$  makes it straightforward to predict the dependence of the
critical force  on the density of crowders.   Our predictions are amenable to experimental
verification using Laser Optical Tweezers.\\
\end{abstract}

\maketitle

\section*{Introduction}

The study of the protein folding problem was galvanized by using concepts from 
the physics of disordered systems. Using a coarse-grained description of folding, expressed in terms of an 
uncorrelated distribution of energies  of protein conformations corresponding to the values at local minima in a multi-dimensional energy landscape, Bryngelson and Wolynes \cite{Bryngelson87PNAS,Bryngelson89JPC} mapped the problem of equilibrium statistical mechanics of protein folding to a random energy model in which the native state plays a special role. These influential works and subsequent studies \cite{Bryngelson95Proteins} showed that most naturally evolved sequences  are foldable, which means that they reach the stable native state on biologically relevant time scales. In this picture, foldable sequences  are characterized by large differences in  the  environmental-dependent folding temperature ($T_f$) and the glass transition temperature ($T_g$) at which the kinetics becomes so sluggish 
that  the folded state is inaccessible on biologically relevant time scales. Related ideas rooted in polymer physics further showed that the interplay of $T_f$, and the equilibrium collapse temperature ($T_{\theta}$) \cite{Camacho93PNAS} could be used to not only fully characterize the phase diagram of  generic protein sequences but also determine their foldability, a prediction that has been experimentally validated very recently \cite{Hofmann12PNAS}. In the intervening years, an impressively large number of important  theoretical and experimental works (for a recent collection see \cite{Wolynes12PNAS} and references cited therein), on a variety of seemingly unrelated problems associated with protein folding have appeared, thus greatly expanding the scope and utility of concepts from statistical mechanics and polymer physics.  Through these developments an expansive view of protein folding and its role in biophysics has emerged \cite{Thirumalai10ARB} with current applications ranging from assisted folding \cite{Betancourt99JMB,Baumketner03JMB,Jewett04PNAS,Hyeon06PNAS}  to describing the functions of molecular motors \cite{Hyeon07PNAS,Hyeon07PNAS2,Tehver10Structure,Koga06PNAS,Zhang12Structure} using models originally devised to understand protein folding kinetics.

A particularly important problem that has benefited 
from the focus on protein folding is the role molecular crowding plays in modulating the thermodynamics and kinetics of folding of proteins \cite{Zhou08ARB} and RNA \cite{Pincus08JACS,Denesyuk11JACS,Kilburn10JACS}  although its importance was recognized long ago \cite{Minton81Biopolym}. It is now widely appreciated that the cytosol is a crowded heterogeneous medium containing a variety of macromolecules such as ribosomes, lipids, and RNA. As a result, folding, diffusion, and other biological processes  in such an environment could be different from what transpires under infinite dilution conditions.  The effects of macromolecular crowding on the stability of synthetic as well as  biopolymers have been  extensively investigated  \cite{WilfBiochemistry1981,Minton81Biopolym,Thirumalai88PRA,MintonARBBS1993,MurphyFEBSLett1996,YoshikawaBiophysChem1999,MintonJMB03,Minton1:2005,Cheung05PNAS,Cheung07JPCB,Cheung07PNAS,Cheung08PNAS,Zhou08ARB,Mittal10BJ,Denesyuk11JACS,Kilburn10JACS} because of the potential relevance  for folding under cellular conditions. In general, several interaction energy and length scales determine whether crowding agents stabilize, have negligible effect, or even destabilize the folded states of proteins \cite{Cheung05PNAS}.  As a result a number of scenarios can emerge depending on the nature of crowding agents, and the choice of proteins. The simplest scenario arises when both the crowder-crowder and crowder-protein interactions are dominated by excluded volume. Although this situation may not accurately characterize even {\it in vitro} experiments it has the advantage that folding in this situation can be  described using a combination of scaling arguments and simulations \cite{Cheung05PNAS}.  Nonspecific athermal crowders (only the excluded volume interactions between the crowders and the crowder and the protein are relevant)
tend to shift the folded$ \rightleftharpoons $unfolded (or equivalently
the zipped$ \rightleftharpoons $unzipped) equilibrium of a biopolymer
towards the folded state by the entropic stabilization theory (EST), because this maximizes the
free-volume available (and hence entropy) to the crowding agents.  This simple theory is based on the  the elegant concept  
of depletion interaction \cite{Asakura:1954,Asakura:1958,Vrij:1976,Verma:1998}, which   posits that the crowding particles decrease the entropy of the unfolded state to a greater extent than the folded state, thus differentially stabilizing the ordered structure \cite{Cheung05PNAS}.  

The EST can be validated by measuring the dependence of the melting temperature on the volume fraction, $(\varphi _{c})$,  of the crowding particles. If EST is valid then the increase, $\Delta T_m(\varphi _{c}) = T_m(\varphi _{c}) - T_m(0)$, should increase. Indeed, absence of any change in $\Delta T_m(\varphi _{c})$ indicates that enthalpic effects play an important role.   Another way to quantify the extent of stabilization
is to ask what critical force $(F_{c})$ would be necessary to unfold a
biopolymer at a given volume fraction $(\varphi _{c})$ of the crowding
agents.  In this paper, we study the simple case of unzipping a polypeptide chain, which forms a $\beta$-hairpin, by applying mechanical force 
as a function of  volume fractions of monodisperse spherical crowding particles.   The study of the zipping/unzipping of biopolymers has a rich history \cite{ZimmJMB1965,KittelAJP1969,FreireBiopolymers1977,AzbelBiopolymers1980,VologodskiiJBioStrucAndDynam1986,RouletPNAS97,BishopPRL1989,Bockelmann:1998,GaubBiophysJ2000,SugimotoNAR2000,Klimov:2002,Hyeon05PNAS}, and has even formed the basis of assessing folding mechanisms of proteins \cite{Munoz97Nature}. 

Surprisingly, there have been very few experimental \cite{Ping:2006,Yuan08ProtSci} or theoretical studies \cite{Pincus08JACS} investigating the effect of mechanical force on proteins  
in a crowded environment.  The experimental studies have argued that $F_{c}$
increases linearly with $\varphi _{c}$ whereas the theoretical arguments \cite{Pincus08JACS} predict a non-linear dependence, which was shown to provide a good fit to the experimental data.  In
this paper, we argue that the unzipping of a biopolymer under {\itshape
constant} tension could be consistent with linear dependence only for small
$\varphi _{c}$.  At higher volume fractions $F_{c}$ does increase non-linearly with $\varphi _{c}$. The increase in $F_{c}$, relative to its value at $\varphi _{c} = 0$, linked to crowding-induced stability, arises because of a 
depletion of the crowding particles from the proximity of proteins. 
This, in turn, results in the crowding agents exerting an osmotic pressure on the biopolymer.  
Unzipping the biopolymer requires that the imposed
tension perform work against  this osmotic  pressure.  Thus,
it is natural to assume that $F_{c}$ should be linearly dependent on the
average pressure $(\mathrm{P})$ associated with crowding particles modeled as
hard spheres.  We have verified this relation using extensive
Monte-Carlo simulations and we present a simple method for determining
$F_{c}$ at an arbitrary $\varphi _{c}$ once the linear dependence
of $F_{c}$ on $\mathrm{P}$ is known.

\section*{Methods}

{\bf Model:} In order to explore the effects of  crowding on the
unzipping of a biopolymer, we chose 
the 16 residue sequence, which forms a $\beta$-hairpin structure, which had been previously used to illustrate the effects
 of confinement on protein folding \cite{Klimov:2002}. The
structure corresponds to the C-terminal $\beta $-hairpin of protein
G (PDB Accession ID 1GB1),  a model system that has been extensively studied using computations \cite{Klimov00PNAS,Pande99PNAS,Dinner99PNAS,Bryant00BJ,Zhou01PNAS,Best11PNAS,Bhattacharya12BJ} following an initial pioneering experimental study \cite{Munoz97Nature}. 

In our simulations, we used a coarse-grained representation of the polypeptide chain. We modeled the hairpin as a collection of 
$N_p=16$ spheres of diameter $\sigma _p=0.38$ nm (each representing a residue)
with configuration $\left\{ {\text{\boldmath $r$}}_{i}\right\}_{i=1}^{N_p}$,
and crowders as a monodisperse collection of $N_c$ spheres of
diameter $\sigma _c=1.0$ nm with configuration ${\{{\text{\boldmath
$R$}}_{I}\}}_{I=1}^{N_{c}}$.  
The Hamiltonian depended on both the positions of
the crowders and conformations of the polypeptide chain:
\begin{multline}
\mathcal{H}( \left\{ {\text{\boldmath $r$}}_{i}\right\} ,\left\{
{\text{\boldmath $R$}}_{I}\right\} ) \equiv \\
\mathcal{H}_{\mathrm{cc}}( \left\{ {\text{\boldmath $R$}}_{I}\right\}
) +\mathcal{H}_{\mathrm{pp}}( \left\{ {\text{\boldmath $r$}}_{i}\right\}
) +\mathcal{H}_{\mathrm{pc}}\left( \left\{ {\text{\boldmath $r$}}_{i}\right\}
,\left\{ {\text{\boldmath $R$}}_{I}\right\} \right) +\mathcal{H}_{\mathrm{bond}}(
\left\{ {\text{\boldmath $r$}}_{i}\right\} ) +\mathcal{H}_{\mathrm{coop}}(
\left\{ {\text{\boldmath $r$}}_{i}\right\} ).
\label{XRef-Equation-76121622}
\end{multline}

\noindent The first three terms on the right hand side (RHS) \ of Eq.\ (\ref{XRef-Equation-76121622}) 
accounted for non-bonded crowder-crowder
(cc), protein-protein (pp), and protein-crowder (pc) interactions
respectively.  The penultimate term on the RHS of Eq.\ (\ref{XRef-Equation-76121622})
$(\mathcal{H}_{\mathrm{bond}}( \{{\text{\boldmath $r$}}_{i}\}) )$
was used to enforce chain connectivity, while the final term on
the r.h.s.\ of Eq.\ (\ref{XRef-Equation-76121622}) ($\mathcal{H}_{\mathrm{coop}}(
\{{\text{\boldmath $r$}}_{i}\}) $) was used to ensure that the hairpin
underwent a cooperative unzipping transition under tension.  The 
interactions between the crowding particles were taken to be,
\begin{equation}
\mathcal{H}_{\mathrm{cc}}( \left\{ {\text{\boldmath $R$}}_{I}\right\}
) =\sum \limits_{J>I}v_{\mathrm{cc}}( \left| {\text{\boldmath $R$}}_{I}-{\text{\boldmath
$R$}}_{J}\right| ) \text{},
\end{equation}

\noindent where
\begin{equation}
v_{\mathrm{cc}}( r) =\begin{cases}
\infty  & \left( r\leq \sigma _{c}\right)  \\
0 & \left( r>\sigma _{c}\right) . \\
\end{cases}
\end{equation}

\noindent Similarly, we used hard-sphere potentials to model the interactions between the 
crowders and the polypeptide (pc):
\begin{equation}
\mathcal{H}_{\mathrm{pc}}( \left\{ {\text{\boldmath $r$}}_{i}\right\}
,\left\{ {\text{\boldmath $R$}}_{I}\right\} ) =\sum \limits_{i,I}v_{\mathrm{pc}}(
\left| {\text{\boldmath $R$}}_{I}-{\text{\boldmath $r$}}_{i}\right|
) \text{},
\end{equation}

\noindent where
\begin{equation}
v_{\mathrm{pc}}( r) =\begin{cases}
\infty  & \left( r\leq \left( \sigma _{p}+\sigma _{c}\right) /2\right)
\\
0 & \left( r>\left( \sigma _{p}+\sigma _{c}\right) /2\right) . \\
\end{cases}
\end{equation}

\noindent The term $\mathcal{H}_{\mathrm{pp}}( \{{\text{\boldmath $r$}}_{i}\},\{{\text{\boldmath
$r$}}_{i}^{0}\}) $ in Eq.\ (\ref{XRef-Equation-76121622}) was decomposed into native ($N$) and non-native
($\mathrm{NN}$) contributions by partitioning the set of residue-residue
distances into those that were less than a cutoff ($R_{\mathrm{cut}}=0.8$ nm)
in the crystal structure and those greater than $R_{\mathrm{cut}}$
(i.e., \{$\{|{\text{\boldmath $r$}}_{i}-{\text{\boldmath $r$}}_{j}|$\}$
\equiv $\{r${}_{\mathrm{ij}}$\}=\{r${}_{\mathrm{ij}}$:$|{\text{\boldmath
$r$}}_{i}^{0}-{\text{\boldmath $r$}}_{j}^{0}|\leq R_{\mathrm{cut}}$\}$
\cup $\{r${}_{\mathrm{ij}}$:$|{\text{\boldmath $r$}}_{i}^{0}-{\text{\boldmath
$r$}}_{j}^{0}|>R_{\mathrm{cut}}$\}).\ \ Letting $\eta =\{r{}_{\mathrm{ij}}:|{\text{\boldmath
$r$}}_{i}^{0}-{\text{\boldmath $r$}}_{j}^{0}|\leq R_{\mathrm{cut}}\}$
and $\vartheta =\{r{}_{\mathrm{ij}}:|{\text{\boldmath $r$}}_{i}^{0}-{\text{\boldmath
$r$}}_{j}^{0}|>R_{\mathrm{cut}}\}$ we write:
\begin{gather}
\mathcal{H}_{\mathrm{pp}}( \left\{ {\text{\boldmath $r$}}_{i}\right\}
,\left\{ {\text{\boldmath $r$}}_{i}^{0}\right\} ) =\mathcal{H}_{\mathrm{pp}}^{N}(
\eta ) +\mathcal{H}_{\mathrm{pp}}^{\mathrm{NN}}( \vartheta ) .
\\\mathcal{H}_{\mathrm{pp}}^{N}( \eta ) =\sum \limits_{d\in \eta
}v_{\mathrm{pp}}^{N}( d) ,%
\label{XRef-Equation-715133743}
\end{gather}

\noindent with 
\begin{equation}
v_{\mathrm{pp}}^{N}( d) =\left\{ \begin{array}{ll}
 \infty  & \left( d/d^{0}<0.8\right)  \\
 -\epsilon  & \left( 0.8\leq d/d^{0}<1.2\right)  \\
 0 & \left( d/d^{0}>1.2\right) 
\end{array},\right. 
\end{equation}

\noindent where $d^{0}$ is the value of $d$ in the crystal structure.

\noindent Similarly,
\begin{equation}
\mathcal{H}_{\mathrm{pp}}^{\mathrm{NN}}( \vartheta ) =\sum \limits_{d\in
\vartheta }v_{\mathrm{pp}}^{\mathrm{NN}}( d) ,
\end{equation}

\noindent where 
\begin{equation}
v_{\mathrm{pp}}^{\mathrm{NN}}( d) =\left\{ \begin{array}{ll}
 \infty  & \left( d\leq \sigma _{p}\right)  \\
 0 & \left( d>\sigma _{p}\right) 
\end{array}.\right. 
\end{equation}

\noindent Chain connectivity was enforced with a sum of box-like
terms:
\begin{equation}
\mathcal{H}_{\mathrm{bond}}( \left\{ {\text{\boldmath $r$}}_{i}\right\}
) =\sum \limits_{i<N_{p}}v_{\mathrm{bond}}( \left| {\text{\boldmath
$r$}}_{i+1}-{\text{\boldmath $r$}}_{i}\right| ) ,
\end{equation}

\noindent where
\begin{equation}
v_{\mathrm{bond}}( r) =\left\{ \begin{array}{ll}
 \infty  & r/r_{b}^{0} < 0.8 \\
 0 & 0.8 \leq r/r_{b}^{0} \leq 1.2 \\
 \infty  & r/r_{b}^{0} > 1.2
\end{array},\right. 
\end{equation}

\noindent and $r_{b}^{0}$ is an ideal $C_{\alpha }-C_{\alpha }$`bonding'
distance of $0.38$ nm.

\noindent The cooperativity term ($\mathcal{H}_{\mathrm{coop}}(
\{{\text{\boldmath $r$}}_{i}\}) $) is a coarse-grained representation
of hydrogen-bonding type interactions and has a nearest-neighbor
Ising-like character,
\begin{equation}
\mathcal{H}_{\mathrm{coop}}( \left\{ {\text{\boldmath $r$}}_{i}\right\}
) =-J\sum \limits_{l=2}^{\mathrm{ncoop}}\Theta ( d_{l}^{0}-d_{l})
\Theta ( d_{l-1}^{0}-d_{l-1}) ,%
\label{XRef-Equation-7615540}
\end{equation}

\noindent where $J=\epsilon /5$, $\Theta ( x) $ is a Heaviside function,
and $d_{l}$ ($d_{l}^{0}$) is the distance (PDB distance) separating
a pair of complementary residues in the strand ($l$ and $l-1$ denote
nearest neighbor pairs).  There were $\mathrm{ncoop}=7$ pairs of
complementary residues in the strand with PDB numbering:\\ 
$\{\{41,56\},\{42,55\},\{43,54\},\{44,53\},\{45,52\},\{46,51\},\{47,50\}\}$ (see Figure~1 for the numbering of the residues as well as the seqence).  Note that not all of these residue pairs are hydrogen bonded in the native
hairpin.  In general, strand pairs exist as parts of larger $\beta
$-sheets and make some hydrogen bonds between the strands of the
pair as well as some hydrogen bonds with other strands of the sheet.  The
coarse-grained nature of Eq.\ (\ref{XRef-Equation-7615540}) renders
the model sufficiently general to ensure transferability to models
of RNA and/or DNA hairpins.  Under such circumstances Eq.\
(\ref{XRef-Equation-7615540}) would mimic the stacking interactions,
which are known to stabilize nucleic acids.

{\bf Simulation Methods.}  
We used a standard Metropolis algorithm  to simulate the
model described by Eq.\ (\ref{XRef-Equation-76121622}) and to
obtain thermodynamic quantities of interest.  Crowder trial moves
were attempted in a `single-spin flip' manner and consisted of random
repositioning of a crowder through the generation of three independent
and uniformly distributed random variables (r.v.'s) on the interval
$[-L/2,L/2]$, where $L=29.7$ nm is the length of a side
of the cubic simulation box.

\noindent The position of residue 1 of the hairpin was held fixed
at the origin throught all simulations (i.e., ${\text{\boldmath
$r$}}_{1}( t) =\text{\boldmath $0$} \; \forall t$).  The remaining
$N_{p}-1$ residue trial moves were randomly selected from a set of
two possibilities.  One type of move corresponded to that used by Baumg\"artner
and Binder \cite{BinderJCP1979} for simulating a freely jointed chain; a random angle
$\gamma $ was chosen from a uniform distribution on $[0,2\pi)$
and an attempt was made to displace residue $i$ by $\gamma $ radians
along the circle perpendicular to the line connecting residues $i-1$
and $i+1$.  For the residue at the free-end of the chain two random
angles $(\beta ,\gamma )$ were chosen and an attempt was made to
move the residue to a new point on the sphere centered at residue
$N_{p}-1$.  The second type of move corresponded to a random change
in the bondlength connecting residue $i$ to residue $i-1$ $(i=2,3,\ldots
,N_{p})$; a uniform r.v. $\aleph $ on $(0.8,1.2)$ was generated
and an attempt was made to map ${\text{\boldmath $r$}}_{i}\mapsto
({\text{\boldmath $r$}}_{i}-{\text{\boldmath $r$}}_{i-1}) \aleph
+{\text{\boldmath $r$}}_{i-1}$.
A trial move from $\mu \rightarrow \nu $ was accepted
with probability ($A( \mu \rightarrow \nu ) $):
\begin{equation}
A( \mu \rightarrow \nu ) =\left\{ \begin{array}{ll}
  e ^{- ( \mathcal{H}_{\nu }-\mathcal{H}_{\mu })/(k_B T) } e ^{(1/k_B T)
F( z_{\nu }-z_{\mu }) } & \left( \mathcal{H}_{\nu }-\mathcal{H}_{\mu
}\right) -F( z_{\nu }-z_{\mu }) >0 \\
 1 & \mathrm{otherwise}
\end{array}. \right.
\end{equation}
where $T$ is the temperature, $k_B$ is Boltzmann's constant, $\mathcal{H}_\nu$ and $\mathcal{H}_\mu$ are as above, $F$ is the constant tension applied 
to the polymer, and $z_\nu$ (resp. $z_\mu$) is the extension in state $\nu$ (resp. $\mu$) of the polymer in the direction of the applied force. 

{\bf Data Collection.} 
Time, measured in Monte-Carlo Steps (MCS), corresponded
to the attempted displacement of ($N_{c}+N_{p}-1$) particles, since
one end of the chain was always held fixed to the origin.  Data
from a trajectory were collected every 1000 MCS.  Figure~2 
reveals that this is significantly longer than the time required
for the RMSD of the crowding agents from an equilibrated initial
state to plateau at all volume fractions except $\varphi _{c}=0.4$.  Even
at $\varphi _{c}=0.4$, the RMSD has increased substantially after
1000 MCS.  We used 0, 5000, 10000, 15000, and 20000 crowders to simulate crowder volume fractions of 0.0, 0.1, 0.2, 0.3, and
0.4 respectively.  For each $\varphi _{c}$, data was collected
at tensions between 0 pN and 40 pN at one pN intervals.  Snapshots
of simulations at each of the non-zero $\varphi _{c}$ and in the
absence of tension are illustrated in Figure~1.  Data
at each force and each $\varphi _{c}$ was collected from multiple
trajectories starting from previously equilibrated configurations
(in turn based on trajectories initiated from random initial crowder
configurations at both high and low-force hairpin configurations).
\section*{Results.} 

{\bf Radial distribution between crowders and the hairpin:} Figure~3 
is a plot of the radial distribution
$(g(r))$ of crowders about the center of mass of the hairpin versus
distance $(r)$ (i.e., $g(r) =V/N_{c}\langle \sum \limits_{I}\delta
( \text{\boldmath $r$}-({\text{\boldmath $R$}}_{I}-{\text{\boldmath
$r$}}_{\mathrm{cm}})) \rangle $).  The maxima in these plots correspond
to the average diameter of the region to which the hairpin finds
itself confined  $(D)$.\ \ Interestingly,
the plots illustrate that the average size of the region is inversely
proportional to the crowder density (i.e., $D\sim \varphi _{c}^{-1}$).\ \ This
suggests that, perhaps, the region in which the hairpin on average is localized is
 aspherical \cite{Honeycutt89JCP}.\ \ If the region were spherical,
we would expect that $D\sim \varphi _{c}^{-1/3}$.

The observation that $D\sim \varphi _{c}^{-1}$ in conjunction with an approximate mapping between crowding and confinement
could be used to obtain the expected scaling of the dependence of the critical force required to the unfold the $\beta$-hairpin, $F_{c}$, on $\varphi _{c}$. Because the confining region is described by a single length, $D$, the EST can be used to identify  the enhancement in the stability of the ordered state with
the loss in entropy of the unfolded state upon  confinement.  Similar scaling approach, using concepts developed in the context of polymer physics, has been used to study confinements effects on biopolymers \cite{Klimov:2002,Mittal08PNAS,Wang09PNAS,Shental-Bechor11JCP}. 
Using this inherently mean-field argument we expect   
\begin{equation}
F_{c} \approx T \Delta S/\Delta x_u^{\ddagger} \sim (R_g/D)^{1/\nu} \frac{k_BT}{\Delta x_u^{\ddagger}} \sim A\varphi _{c}^{1/\nu}. 
\end{equation}
In the above equation $T \Delta S$ is the penalty for confining the polypeptide chain with dimension $R_g$ in a region with size $D$, $\Delta x_u^{\ddagger}$ is the minimum extension needed to unfold the protein, and $\nu$ is the Flory exponent. Because $D\sim \varphi _{c}^{-1}$ we expect that $F_{c} \approx \varphi _{c}^{1/\nu} \sim \varphi _{c}^{5/3}$ assuming that $\nu \approx 0.6$. If $D\sim \varphi _{c}^{-1/3}$, as would be the case if the unfolded state were spherical, then it follows that  $F_{c} \approx \varphi _{c}^{5/9}$, a result that we derived previously \cite{Pincus08JACS} to analyze the experimental data on forced-unfolding of ubiquitin.

{\bf Numerical evidence for Eq. (15):}  Plots of the average extension of the hairpin $(\langle z \rangle )$ 
versus applied tension $(F)$ presented in Figure~4 {\bfseries (a)} 
show that the $\langle z \rangle $ 
decreases monotonically with $\varphi _{c}$ at moderate
values of $F$.  This implies that crowding in essence decreases $\langle z \rangle $ because the entropic
 penalty to stretch a protein in a crowded environment is far too large. In other words, the probability of finding a region free of crowders decreases exponentially as the extension increases, which explains the observed results in Figure~4 {\bfseries (a)}.  The isothermal extensibility $(\chi \equiv
\partial \langle z\rangle /\partial F)$ plots in Figure~4{\bfseries
(b)} reveal that $F_{c}$ (i.e., the value of $F$ at  which 
$\chi $ is a maximum) increases monotonically with increasing $\varphi _{c}$. 
A plot of $F_{c}$ versus $\varphi _{c}$ (Figure~5{\bfseries
(a)}) subsequently revealed that the power-law dependence of $F_{c}$ on
$\varphi _{c}$ is characterized by an exponent $\alpha \cong 1.6$, which is in accord with the scaling predictions in Eq. (15). 
 Data collapse of $\chi$ based on a scaling function
$X( (F-F_{c})/F_{c})$ that is independent of $N_{c}$ revealed that
$\chi \sim (1-A N_{c}^{d_{\chi}})$, where $d_{\chi} \cong 1.43$ and
$A \cong 1.7 \times {10}^{-7}$.  This shows that the effects of crowding and force can be separated, which to some extent justifies the scaling theory predictions. Thus, when measured in terms of
the reduced distance to the critical force, the primary effect of
the crowders is to decrease the extensibility of the chain.

We can also obtain the dependence of $F_{c}$ on $\varphi _{c}$ using the $F$-dependent changes in an order parameter that characterizes the folded state. The  extent of structure formation can be inferred using the average fraction of native contacts, $\langle Q\rangle $. In Figure~6  we show 
$\langle Q\rangle $ as a function of $F$. For all values of $F$ the crowding particles increase $\langle Q\rangle $, which is a reflection of the enhanced stabilization of the native state of $\beta$-hairpin at  $\varphi _{c} \ne 0$. Let us define $F_m$ using $\langle Q\rangle  = 0.5 $ at $\varphi _{c} = 0$. At this value of $F_m$, Figure~6{\bf a} shows that $\langle Q\rangle \approx 0.75$ at $\varphi _{c} = 0.4$.  The critical force $F_{c}$ can identified with the force at which $|\frac{d \langle Q\rangle}{dF}|$ (Figure~6{\bf b}) achieves a maximum.  
It is clear that $F_c$ is an increasing function of $\varphi _{c}$ (Figure~6{\bf c}). Just as in Figure~4{\bfseries
(a)}, where $F_c$ is identified with the maximum in the isothermal extensibility, we find that $F_c \sim \varphi _{c}^{\alpha}$ with $\alpha \approx 1.6$ (Figure~6{\bf c}). The numerical simulations using different measures confirm the scaling predictions showing the power law increase in the $\varphi _{c}$-dependent critical force required to rupture the hairpin.   

{\bf Osmotic (or disjoining) pressure explains the origin of $\varphi _{c}$-dependent $F_{c}$:}   
Insights into our results can be obtained by viewing the depletion forces from a 
different perspective.  Because of the repulsive interaction between the crowders and the polypeptide chain the crowding particles are depleted from the surface of the protein. In the process, the crowding particles not only gain translational entropy but they also exert an osmotic pressure on the polypeptide chain, thus forcing it to adopt a compact structure. In other words, the crowders can be viewed as providing
an isothermal and {\itshape isobaric} bath for the hairpin.  In such a case, 
it is natural to assume that 
$F_{c}$ is proportional to the average
pressure $(\mathrm{P})$ associated with a hard sphere fluid at that
density and volume fraction: 
\begin{equation}
F_{c}=m \mathrm{P}+b.
\end{equation}
where $m$ and $b$ are constants to be determined.

The disjoining or osmotic pressure can, in turn, be calculated from the contact
value $g( \sigma ) \equiv \operatorname*{\lim }\limits_{r\,\rightarrow
\:\sigma ^{+}}g( r)$ of the crowder-crowder radial distribution
function using the standard relation, 
\begin{equation}
\mathrm{P}=\rho  k_{B}T( 1+4\varphi _{c}g( \sigma ) ) .%
\label{XRef-Equation-73110918}
\end{equation}

\noindent Equation (\ref{XRef-Equation-73110918}) follows from
the viral-based expression for hard sphere systems,
\begin{equation}
\frac{\mathrm{P}}{\rho  k_{B}T}=1-\frac{2 \pi  \rho }{3 k_{B}T}\int
_{0}^{\infty }g( r) \frac{dv}{dr} r^{3}dr
\end{equation}

\noindent via the substitution $g( r) =\psi ( r)  e^{-\beta  v(
r) }$ and by noting that the Boltzmann factor $e^{-\beta  v}=\theta
( r-\sigma ) $ for hard spheres where $\theta(x)$ is the step function.

From the Figure~7{\bfseries (a)}
showing the crowder-crowder $g( r) $ at volume fractions $\varphi
_{c}=$ 0.1, 0.2, 0.3, and 0.4 we computed $g( \sigma )$, which  was subsequently
used to determine the average pressure at each $\varphi _{c}$.  
The linear correlation coefficient $(r)$ between the two variables ($F$ and $\mathrm{P}$)
was determined to be 0.99657 (Figure~7{\bfseries
(b)}).  The probability that 5 measurements of two uncorrelated
random variables would yield a correlation coefficient this high
is $\frac{2\Gamma ( 2) }{\sqrt{\pi }\Gamma ( 3/2) }\int _{0.99657}^{1}{(1-x^{2})}^{1/2}dx=0.00024$,
where $\Gamma ( x) $ is Euler's gamma function.  In
 Figure~7{\bfseries (b)} we provide the
best fit line to the data yielding $m=0.24$ ${\mathrm{nm}}^{2}$ and
$b=13.84$ pN. Thus, $F_{c}$ is linearly related to the osmotic pressure arising from depletion forces, whose strength
is a  measure of the stabilization of the ordered state.

\noindent  In order to obtain the dependence of $F_{c}$ on $\varphi _{c}$ a reliable relationship between $\mathrm{P}$ and 
$\varphi_c$ needs to be established. Although $\mathrm{P}$ can be calculated using simulations it would be convenient to obtain approximate analytically calculable estimates of $\mathrm{P}$. The average pressure associated with the hard spheres
can be determined at all $\varphi _{c}$ using the successful semi-empirical Carnahan-Starling
equation of state:
\begin{equation}
\frac{\mathrm{P}}{\rho  k_{B}T}=\frac{1+\varphi _{c}+\varphi _{c}^{2}-\varphi
_{c}^{3}}{{\left( 1-\varphi _{c}\right) }^{3}}.%
\label{XRef-Equation-73111718}
\end{equation}

\noindent Figure~7{\bfseries (c)} shows $F_{c}$ versus
$\varphi _{c}$ (blue circles) as well as the curve associated with
the best-fit linear relation between $F_{c}$ and $\mathrm{P}$ 
calculated using Eq.\ (\ref{XRef-Equation-73111718}).  This
linear relation yielded $m=$ 0.17 nm{}\textsuperscript{2}
and $b=13.96$ pN, which are close to the values obtained by fitting $F_c$ to numerically computed values for $\mathrm{P}$.  The stars illustrate $(\varphi _{c},F_{c})$
ordered pairs associated with the best-fit linear relation between
$F_{c}$ and $\mathrm{P}$ as calculated from Eq.\ (\ref{XRef-Equation-73110918}) (as in Figure~7{\bfseries (b)}). We surmise that  using $\mathrm{P}$  approximated by Eq.\ (\ref{XRef-Equation-73111718}) can be used to obtain accurate estimates of $F_{c}$ given knowledge of the coefficients m and b. 

Finally, the accuracy of the Carnahan-Starling equation is assessed by plotting in 
Figure~7{\bfseries
(d)}  Eq.\ (\ref{XRef-Equation-73111718}) as well as several
truncated Taylor-series expansions:
\begin{equation}
\mathrm{P}\cong \frac{6k_{B}T}{{\pi \sigma }^{3}}\sum \limits_{i=1}^{n}a_{i}\varphi
_{c}^{i},
\end{equation}

\noindent of this equation of state. Note
that the relative error associated with the linear approximation
$(n=1)$  of 0.342651 is large at $\varphi _{c}=0.1$, while the relative
error is only 0.0326804 at $\varphi _{c}=0.4$ when $n=7$ terms are
included in the expansion.  The coefficients of the expansions
are $a_{1}=1,a_{2}=4,a_{3}=10,a_{4}=18,a_{5}=28,a_{6}=40,$ and $a_{7}=54$.  The linear relation between $F_{c}$ and $\mathrm{P}$ shows that, close to $\varphi _{c} = 0$, $F_{c}$ should depend only linearly on $\varphi _{c}$ because $\mathrm{P}$ is approximately linearly dependent on $\varphi _{c}$ for small $\varphi _{c}$.
However, in order to
determine the value of $F_{c}$ at an arbitrary $\varphi _{c}$, one
should first determine the appropriate linear relation between $F_{c}$
and $\mathrm{P}$.  The critical force $F_{c}$ can then be determined at an arbitrary
$\varphi _{c}$ using the  linear relation and approximate estimate of $\mathrm{P}$ given in Eq.\ (\ref{XRef-Equation-73111718}). 

\section*{Conclusions.}

Using simple theoretical arguments and extensive MC simulations
of a three dimensional off-lattice model, we have demonstrated for the
first time that the critical force for unzipping a biopolymer under
tension obeys a non-linear dependence on the volume fraction of
crowding agent.  This dependence can be characterized by a power
law dependence with an exponent $\alpha \cong 1.6$: $F_{c} \sim \varphi
_{c}^{\alpha}$. The exponent $\alpha$ is surprisingly close to the scaling prediction $1/\nu$ with $\nu \approx 3/5$. 

The numerical findings and scaling predictions can be understood by noting that the
crowders provide an isobaric environment for the protein.  The osmotic pressure arises from the depletion forces due to
expulsion of the crowding particles from the protein, and is entropic in origin. Because of the osmotic pressure  unzipping
requires that the tension imposed on the hairpin perform mechanical work
against the isotropic pressure.  These arguments are fully confirmed in simulations, which demonstrate that $F_{c}$ has a highly  significant
linear correlation with the pressure $(\mathrm{P}\mathrm{)}$
of the hard-sphere crowding particle in which it is embedded.  To determine $F_{c}$ at an arbitrary
$\varphi _{c}$, one should first determine the linear dependence
of $F_{c}$ on $\mathrm{P}$.  The exact relation connecting $F_{c}$
to $\varphi _{c}$ then follows from the Carnahan-Starling equation
of state.  This relationship shows that  $F_{c}$  displays an approximately linear dependence
on $\varphi _{c}$ for volume fractions near $\varphi _{c}=0$. However, when examined over a large range of $\varphi_c$ we expect that $F_{c}$ should increase non-linearly with 
$\varphi_c$ as indicated by the scaling predictions exploiting the relationship between crowding and confinement.  

Two comments about the scaling predictions are important to make. (i) The exponent $\alpha$ relating the increase in $F_{c}$ to $\varphi _{c}$, although related to the Flory exponent ($\nu$), is likely to depend both on the nature of the unfolded states of the protein and the shape of the crowding particles. If the overall shape of the unfolded state is non-spherical as is clearly the case for the hairpin (Figure 1) then $\alpha \approx$ 1.6. On the other hand, if the unfolded state is spherical on an average, as is likely to be the case for larger proteins, then  it is likely $\alpha \approx 5/9$, as argued previously \cite{Pincus08JACS}.  
(ii) The theoretical predictions are based on a mean-field picture in which it is assumed that crowding 
(modeled with hard spheres) results in the protein being localized to
 a cavity.  Thus, fluctuations in the crowding particles are ignored. These, especially close to the protein, could have significant effects. The good agreement between scaling predictions and simulations suggests that the fluctuation effects are not significant, at least for the case tested here. In principle, the importance of fluctuations can be tested by fixing the locations of the crowding particles. Such quenched simulations are equivalent to the present annealed simulations for the properties of the proteins because in a large sample containing fixed obstacles the protein would sample many distinct environments. This is then the same as performing annealed simulations. Therefore, we expect that the scaling properties predicted and tested here will not change even if the simulations are done by fixing the locations of the crowding particles.  Additional simulations on proteins, rather than polypeptide chains forming secondary structures,  would be needed to obtain accurate values of $\alpha$.

There are only very few experiments probing the limits of mechanical stability of proteins in the presence of crowding agents.  For
example,  Atomic Force Microscopy has been used to  investigate
the effects of dextran on the mechanical stability
of proteins \cite{Yuan08ProtSci}.  These researchers found
$F_{c}\sim \varphi _{c}$ for $\varphi _{c}\in [0.0,0.3]$ and non-linearity
only for $\varphi _{c}>0.3$.  It is important to note 
that their experimental setup is of an inherently non-equilibrium
character; one end of a protein is extended at a constant speed
while the other end is used to probe the chain's tension.  Furthermore, the protein
examined (Ubiquitin) is unlikely to be a two-state folder  and may undergo distinct
unzipping reactions at multiple tensions. In this case, the effects of crowding in an energy landscape with multiple barriers \cite{Hyeon12JCP} may have to be studied.  Because of the non-equlibrium nature of the AFM setup it would be desirable to verify the predictions of the present work using laser optical tweezer experiments in which small constant forces can be applied. 

{\bf Acknowledgements:} We are grateful to the National Institutes of Health (GM089685) for supporting the research. The
authors kindly thank the National Energy Research Scientific Computing
(NERSC) Center for significant computational time and resources.

\newpage

\bibliography{./ForceAndCrowding.master.bib}
\bibliographystyle{pnas}

\newpage
\begin{center}
\textbf{\large{Figure Captions}}
\end{center}

\hspace{-1.2em}Figure~1: Snapshots of the simulated system in the absence of mechanical force.
 {\bfseries (a) }$\varphi _{c}=0.1$ ($N_{c}=5000$), {\bfseries
(b)} $\varphi _{c}=0.2$ ($N_{c}=10000$), {\bfseries (c)} $\varphi
_{c}=0.3$ ($N_{c}=15000$), and {\bfseries (d)} $\varphi _{c}=0.4$ ($N_{c}=20000$).  The
hairpin corresponds to the small dark spot at the center of each
of the boxes. The center of the figure shows a blowup of the region adjacent to the hairpin. The purpose of showing the four snapshots is to illustrate that the biopolymer is jammed in a sea of crowding particles. The blowup in the center shows structure of the $\beta$-hairpin along  with the sequence numbering of the 16 residues. 

\bigskip

\hspace{-1.2em}Figure~2: Root Mean Square Deviation (RMSD) of the crowding agents
from an equilibrated initial state as a function of time $(\tau
$ - measured in Monte Carlo Steps per free-particle (MCS)) for trajectories
at $\varphi _{c}=0.1$ (blue), $\varphi _{c}=0.2$ (green), $\varphi
_{c}=0.3$ (orange), and $\varphi _{c}=0.4$ (red).  Note that crowder
trial moves are reasonably successful for $\varphi _{c}\leq 0.3$, which implies that the allowed conformations are ergodically sampled.
The acceptance ratio for such trial moves is significantly
reduced at $\varphi _{c}=0.4$ although the errors in the results are small as indicated by consistency between different measures.  The sampling interval used for
collecting the data presented in all figures below was $1000$ MCS.

\bigskip

\hspace{-1.2em}Figure~3: Radial distribution function $(g(r))$ of the crowders
about the center of mass of the hairpin at various volume fractions
and at $F=0$ pN.  Red squares, orange diamonds, green upward triangles, and blue downward triangles respectively correspond to $\varphi _{c} = 0.1, 0.2, 0.3,$ and 0.4.  The  
maxima in $g(r)$ lie at $r=15.5, 14.5, 13.5,$ and $12.5$ {\AA} respectively.  This
suggests that the size of the
region to which the hairpin is confined $(D)$ is inversely proportional
to $\varphi _{c}$, implying that this region is aspherical.   The symbols represent raw data, curves correspond to smoothing using Eq.\ (6.48) of Allen and Tildesley \cite{Allen:1987}.

\bigskip

\hspace{-1.2em}Figure~4: Plot of {\bfseries (a)} average extension $(\langle z\rangle
)$ as a function of force $(F)$.  {\bfseries (b)} Isothermal extensibility ${(\chi 
\equiv  \partial \langle z\rangle /\partial F|}_{T}$) versus
force $(F)$.  Black, red, orange, green, and blue curves respectively
correspond to volume fractions $(\varphi _{c})$ of $0,0.1,0.2,0.3,$ and
$0.4$.  All curves were calculated using the multiple histogram reweighting method;
symbols in {\bfseries (a)} correspond to unreweighted data.

\bigskip

\hspace{-1.2em}Figure~5: {\bfseries (a)} Critical force $(F_{c})$ of the hairpin
versus crowder volume fraction $(\varphi _{c})$.  $F_{c}$ displays
a power law dependence on the volume fraction of crowding agent
with an exponent $(\alpha )$ of 1.55.  {\bfseries (b)}  Data
collapse of the isothermal extensibility $(\chi )$ (Figure~4)
shows that $\chi \sim (1-A N_{c}^{d_{\chi }})X( (F-F_{c})/F_{c}) $,
where the scaling function $(X( x) )$ is independent of the number
of crowders $(N_{c})$.\ \ Thus, the dependence of the isothermal
extensibility on $N_{c}$ is characterized by an exponent $(d_{\chi
})$ of 1.43.  Black, red, orange, green, and blue curves respectively
are for $N_{c}=$$0,5000,10000,15000,$ and $20000$.

\bigskip

\hspace{-1.2em}Figure~6: {\bfseries (a)} Average fraction of native contacts
($\langle Q\rangle $) versus applied tension ($F$).  {\bfseries
(b)} Absolute value of $d\langle Q\rangle /dF$ versus $F$.  The
force which maximizes $|d\langle Q\rangle /dF |$ at a particular
volume fraction $\varphi _{c}$ corresponds to the critical force
$F_{c}( \varphi _{c}) $ at $\varphi _{c}$.  {\bfseries (c)} A
plot of $F_{c}$ versus $\varphi$${}_{c}$ verifies the results illustrated
in Figure~5 {\bfseries (a)}; $F_{c} \sim  \varphi
_{c}^{\alpha }$ where $\alpha \cong 1.6$.

\bigskip

\hspace{-1.2em}Figure~7: {\bfseries (a)} Crowder-crowder radial distribution
fucntion $(g( r) )$ versus separation distance ($r/\sigma $).  Red
squares, orange diamonds, green up triangles, and blue down triangles
respectively correspond to $\varphi _{c}=0.1, 0.2, 0.3,$ and 0.4.  {\bfseries
(b)} The contact value $g( \sigma ) \equiv \operatorname*{\lim
}\limits_{r\,\rightarrow \:\sigma ^{+}}g( r) $ from {\bfseries (a)} was
used to calculate the average pressure ($\mathrm{P}$) of the hard
spheres at each $\varphi _{c}$ using the virial derived equation $\mathrm{P}=\rho
k_{B}T( 1+4 \varphi _{c} g( \sigma ) ) $.  A plot of $F_{c}$ versus
$\mathrm{P}$ revealed a highly significant correlation with a linear
correlation coefficient $r=0.99657$.  Blue circles correspond
to measured data and the black solid line corresponds the the best
fit line $F_{c}=m \mathrm{P}+b$, where $m=0.24$ ${\mathrm{nm}}^{2}$
and $b=13.84$ $\mathrm{pN}$.  {\bfseries (c)} The dependence  of $F_{c}$ on
$\varphi _{c}$.  Blue circles correspond to simulation data.  The
solid black curve corresponds to the best-fit line relating $F_{c}$ to
the pressure $(\mathrm{P}( \varphi _{c}) )$ at $\varphi _{c}$, where $\mathrm{P}$
was calculated from the semi-empirical Carnahan-Starling equation
of state: $\mathrm{P}=\rho  k_{B}T\frac{(1+\varphi _{c}+\varphi
_{c}^{2}-\varphi _{c}^{3})}{{(1-\varphi _{c})}^{3}}$.  The red
stars also correspond to a best-fit line relating $F_{c}$ to $\mathrm{P}(
\varphi _{c})$, where P was calculated as in {\bfseries (b)}.  {\bfseries
(d)} P versus $\varphi _{c}$ as calculated via the Carnahan-Starling
equation of state (black) and after truncating a Taylor-series expansion
($\mathrm{P} \cong \frac{6 k_{B}T}{{\pi \sigma }^{3}}\sum \limits_{i=1}^{n}a_{i}\varphi
_{c}^{i}$) of this equation of state about $\varphi _{c}=0$ after
$n=1$(red), 2 (orange), 3 (yellow), 4 (green), 5 (cyan), 6 (blue),
and 7 (purple) terms.

\newpage
\begin{figure}[h]
\begin{center}
\includegraphics[angle=90,width=120mm]{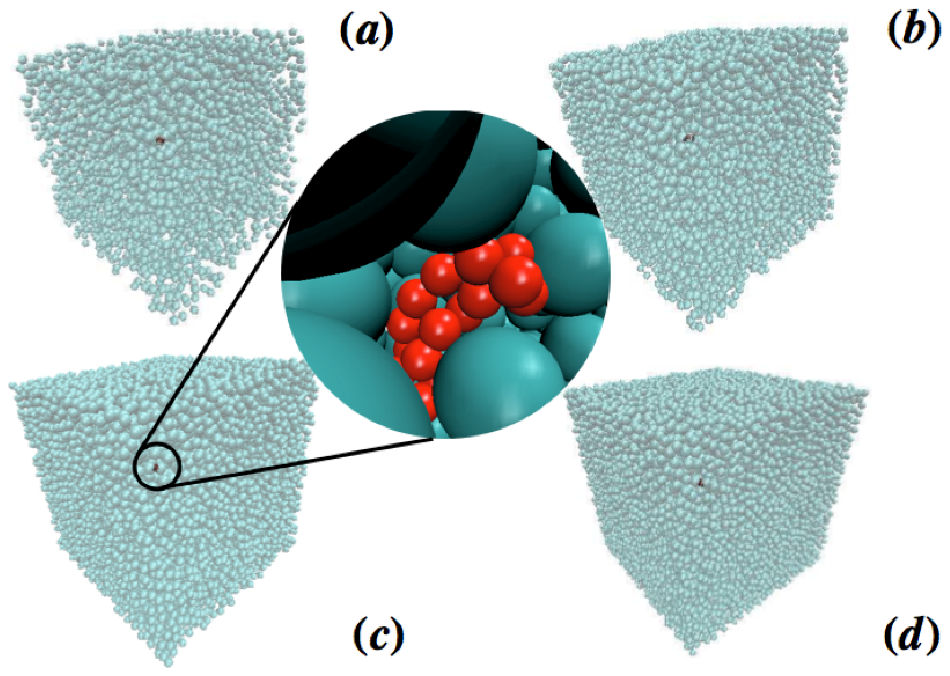}
\end{center}
\end{figure}

\newpage
\begin{figure}[h]
\begin{center}
\includegraphics[width=150mm]{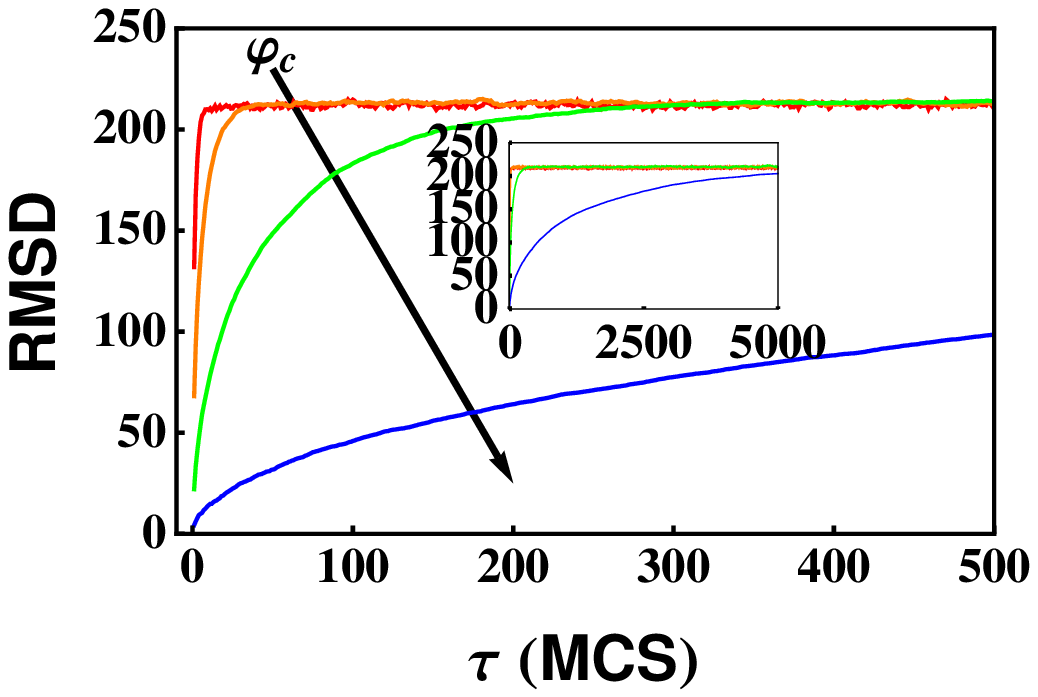}
\end{center}
\end{figure}

\newpage
\begin{figure}[h]
\begin{center}
\includegraphics[width=170mm]{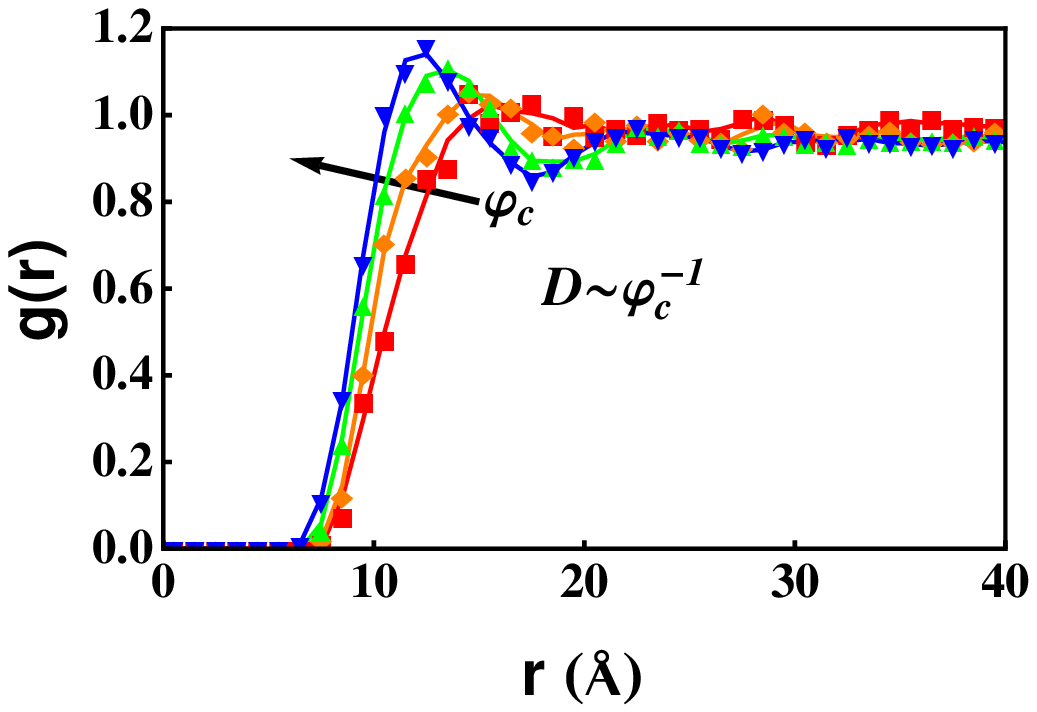}
\end{center}
\end{figure}

\newpage
\begin{figure}[h]
\begin{center}
\includegraphics[angle=90,width=80mm]{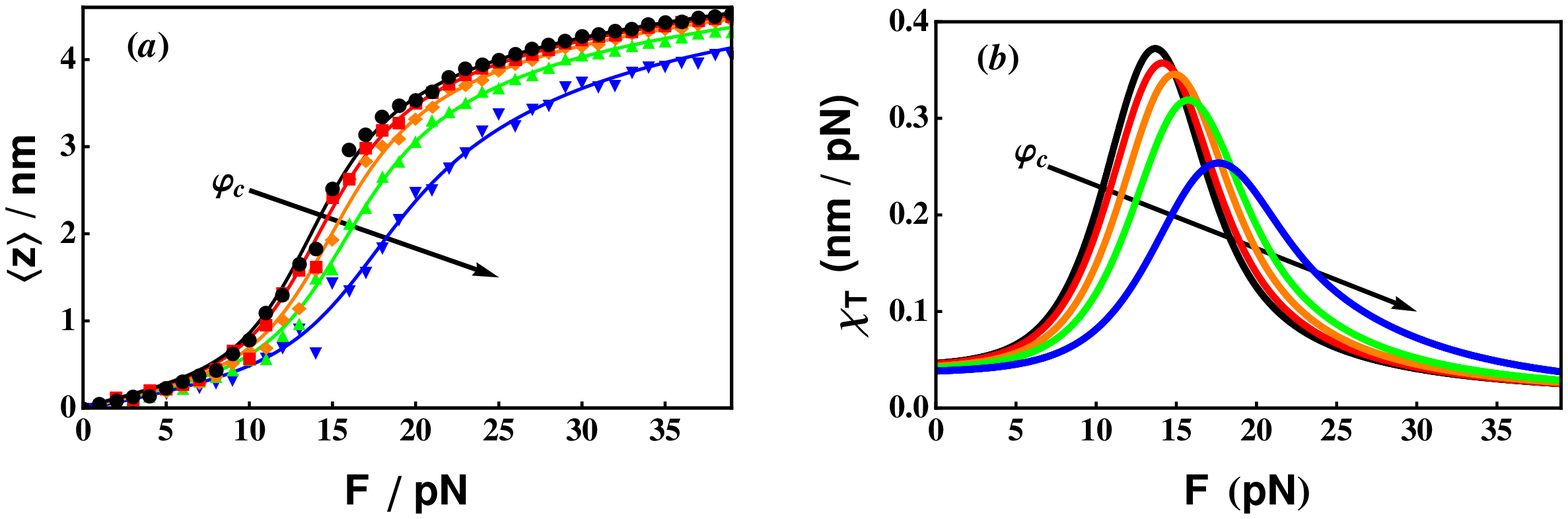}
\end{center}
\end{figure}

\newpage
\begin{figure}[h]
\begin{center}
\includegraphics[angle=90,width=80mm]{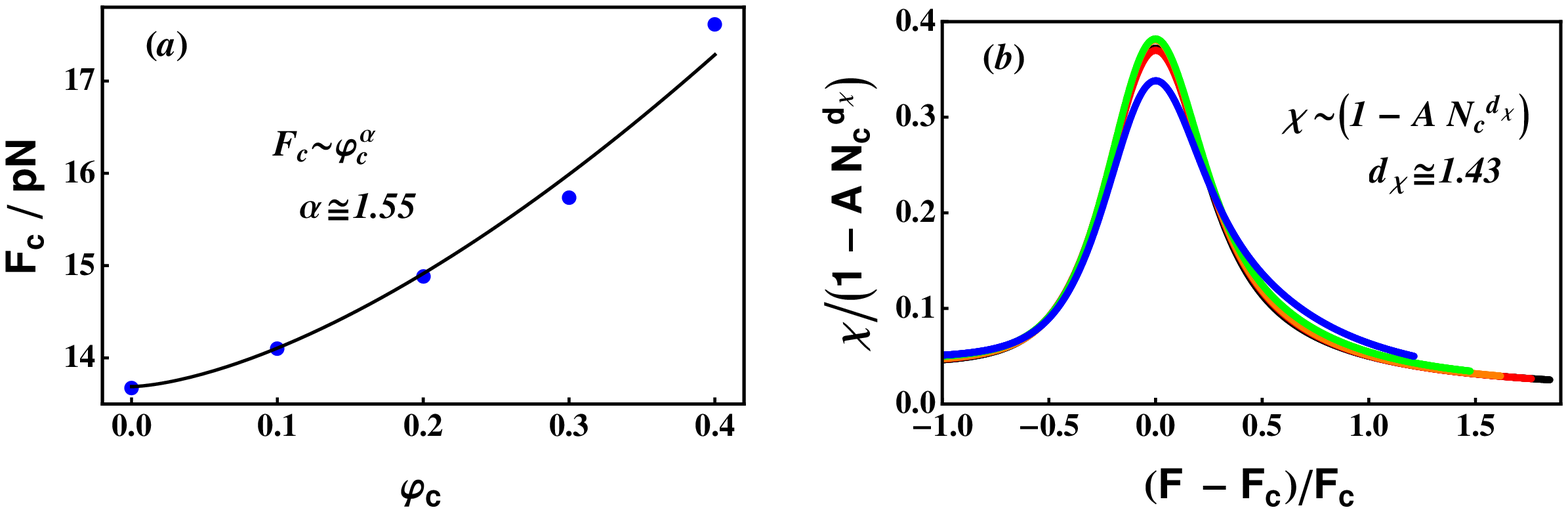}
\end{center}
\end{figure}

\newpage
\begin{figure}[h]
\begin{center}
\includegraphics[width=140mm]{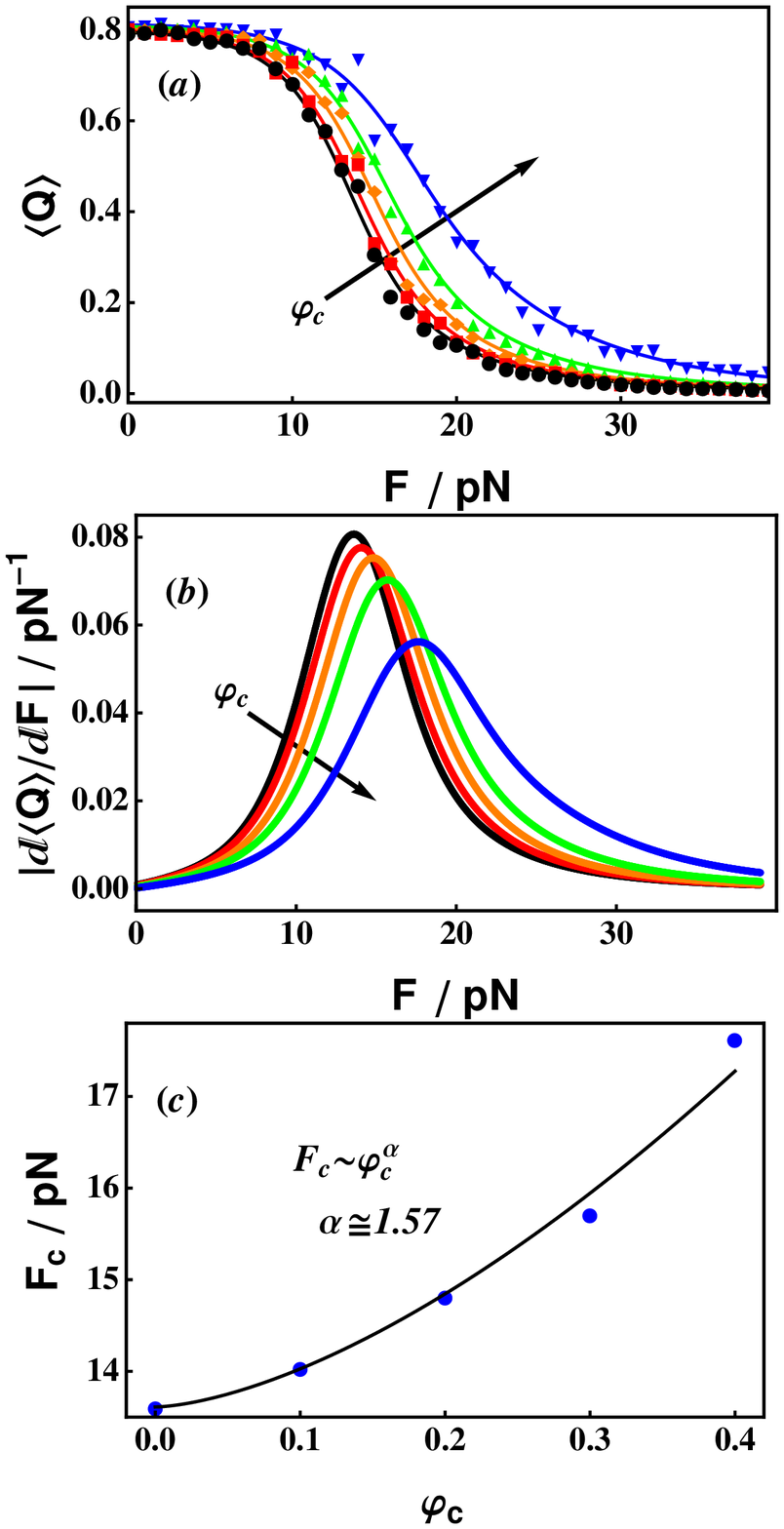}
\end{center}
\end{figure}

\newpage
\begin{figure}[h]
\begin{center}
\includegraphics[angle=90,width=150mm]{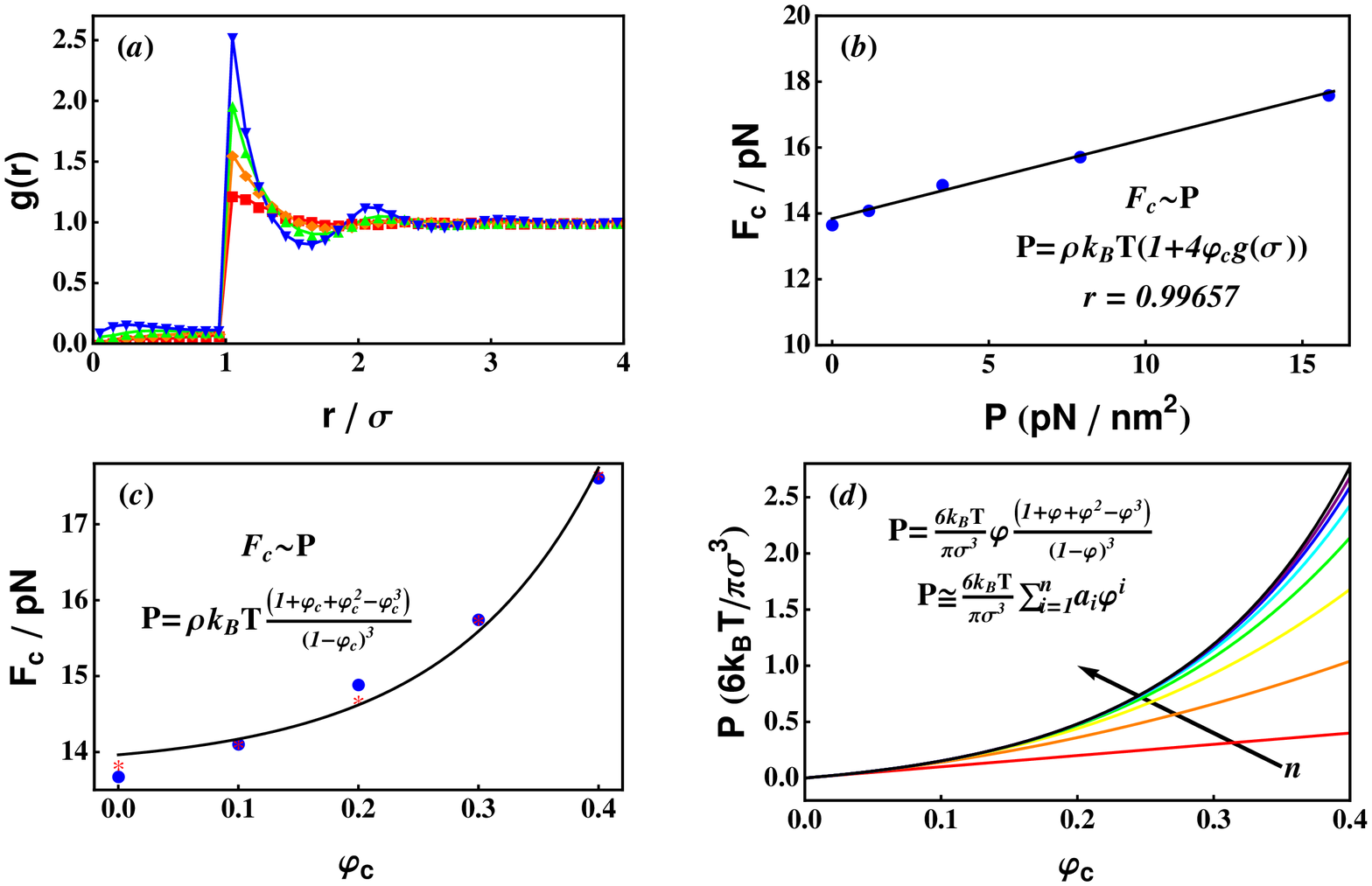}
\end{center}
\end{figure}

\end{document}